\documentclass{article}

\usepackage[nonatbib, preprint]{cig}

\usepackage[utf8]{inputenc} %
\usepackage[T1]{fontenc}    %
\usepackage{hyperref}       %
\usepackage{arydshln}
\usepackage[export]{adjustbox}
\usepackage{tikz}
\usepackage{cite}

\usepackage{cite}
\usepackage{amsmath,amssymb,amsfonts}
\usepackage{algorithmic}
\usepackage{graphicx}
\usepackage{textcomp}
\usepackage{xcolor}

 \usepackage{multirow} 
\usepackage{bm}
\usepackage{algorithm}
\usepackage{tikz}
\usetikzlibrary{positioning, arrows.meta, calc}

\definecolor{darkblue}{rgb}{0, 0, 0.85}
\definecolor{lightgreen}{rgb}{.85,1,.85}
\definecolor{lightred}{rgb}{1,.85,.85}
\definecolor{lightblue}{rgb}{.85,.85,1}
\definecolor{pink}{HTML}{EB346F}
\hypersetup{
    colorlinks = true,
    citecolor = darkblue,
    linkcolor = {black}
}

\def\R{\mathbb{R}}

\def\kbm{{\bm{k}}}

\def\xbm{{\bm{x}}}
\def\zbm{{\bm{z}}}
\def\ybm{{\bm{y}}}
\def\zbm{{\bm{z}}}
\def\sbm{{\bm{s}}}

\def\dbm{{\bm{d}}}
\def\nbm{{\bm{n}}}

\def\Abm{{\bm{A}}}
\def\Dbm{{\bm{D}}}

\def\Ibm{{\bm{I}}}

\def\Abm{{\bm{A}}}
\def\Dbm{{\bm{D}}}

\def\Ibm{{\bm{I}}}

\def\Rcal{{\mathcal{R}}}

\def\Ncal{{\mathcal{N}}}

\def\Dcal{{\mathcal{D}}}

\def\Tsf{{\mathsf{T}}}
\def\Tsf{{\mathsf{T}}}

\def\argmin{\mathop{\mathrm{arg\,min}}} %

\def\gradDen{\Dcal_{\theta}}

\title{Analysis Plug-and-Play Methods for\\Imaging Inverse Problems
}

\vspace{5em}
\author{
    Edward P. Chandler${^1}$ \quad
    Shirin Shoushtari${^1}$ \quad
    Brendt Wohlberg${^2}$ \quad  
    Ulugbek S. Kamilov${^1}$ \\
    [0.7em]
    \small \textnormal{${^1}$Washington University in St.\ Louis} \qquad 
    \small \textnormal{${^2}$Los Alamos National Laboratory}\\[0.5em]
\footnotesize \texttt{\{e.p.chandler, s.shirin, kamilov\}@wustl.edu, brendt@ieee.com
}
}

\begin{document}

\maketitle

\begin{abstract}
    Plug-and-Play Priors (PnP) is a popular framework for solving imaging inverse problems by integrating learned priors in the form of denoisers trained to remove Gaussian noise from images. In standard PnP methods, the denoiser is applied directly in the image domain, serving as an implicit prior on natural images. This paper considers an alternative \emph{analysis} formulation of PnP, in which the prior is imposed on a transformed representation of the image, such as its gradient. Specifically, we train a Gaussian denoiser to operate in the gradient domain, rather than on the image itself.
    Conceptually, this is an extension of total variation (TV) regularization to learned TV regularization.
    To incorporate this gradient-domain prior in image reconstruction algorithms, we develop two analysis PnP algorithms based on half-quadratic splitting (APnP-HQS) and the alternating direction method of multipliers (APnP-ADMM). We evaluate our approach on image deblurring and super-resolution, demonstrating that the analysis formulation achieves performance comparable to image-domain PnP algorithms.
\end{abstract}

\section{Introduction}
\label{sec:intro}
Imaging inverse problems seek to estimate an image from a set of noisy measurements.
A common strategy is to formulate an optimization problem that incorporates both a forward measurement model and a prior that encodes assumptions about the structure of the desired solution. 
In classical signal processing, handcrafted priors such as total variation (TV) have been widely used to regularize inverse problmes~\cite{rudin.etal1992}.
While these hand-crafted priors can work well, they often introduce reconstruction artifacts, such as staircasing in TV-regularized solutions~\cite{strong2003edge}. This limitation arises because handcrafted priors are typically too simplistic to accurately model the complex statistics of natural images.

As deep learning image priors have demonstrated superior capabilities in capturing high-dimensional data distributions, learned priors have become central to state-of-the-art solutions for imaging inverse problems~\cite{bora2017compressed, kadkhodaie2020solving}.
Plug-and-play (PnP) methods have been extensively used to incorporate such learned priors, specified by denoisers, for solving inverse problems~\cite{venkatakrishnan.etal2013, Kamilov.etal2023}. PnP methods are not only empirically successful but also supported by theoretical analyses that guarantee convergence under certain conditions~\cite{liu.etal2020}.

\begin{figure*}[ht]
    \centering
    \begin{tikzpicture}[
        node distance=0.5cm and 0.5cm,
        every node/.style={align=center},
        img/.style={rectangle},
        block/.style={rectangle, draw, rounded corners, minimum width=2cm, minimum height=1.5cm, fill=blue!10},
        arrow/.style={->, thick},
    ]
        
    \node[img] (image_pic) {\includegraphics[width=2.0cm]{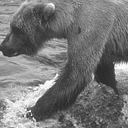}};
    \node[below=0pt of image_pic] {\footnotesize Image};
    
    \node[img, right=of image_pic] (grad_pic) {
        \begin{tikzpicture}[baseline]
            \node[inner sep=0pt] at (0,0) {\includegraphics[width=2.0cm]{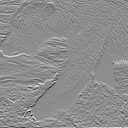}};
            \node[inner sep=0pt] at (0.35,-0.35) {\includegraphics[width=2.0cm]{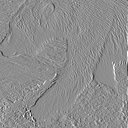}};
        \end{tikzpicture}
    };
    \node[below=0pt of grad_pic] {\footnotesize Image Gradient};
    
    \node[img, right=of grad_pic] (noisy_pic) {
        \begin{tikzpicture}[baseline]
            \node[draw, thick, minimum width=2.0cm, minimum height=2.0cm] at (-0.35,0.35) {\begin{minipage}[t][2.0cm][t]{2.0cm}
                \centering
                \small Noise Level
            \end{minipage}};
            \node[inner sep=0pt] at (0,0) {\includegraphics[width=2.0cm]{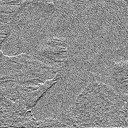}};
            \node[inner sep=0pt] at (0.35,-0.35) {\includegraphics[width=2.0cm]{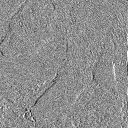}};
        \end{tikzpicture}
    };
    \node[below=0pt of noisy_pic] {\footnotesize Noisy Image Gradient};
    
    \node[block, right=of noisy_pic] (network) {Network};
    
    \node[img, right=of network] (output_pic) {
        \begin{tikzpicture}[baseline]
            \node[inner sep=0pt] at (0,0) {\includegraphics[width=2.0cm]{img/img_grad_0.png}};
            \node[inner sep=0pt] at (0.35,-0.35) {\includegraphics[width=2.0cm]{img/img_grad_1.png}};
        \end{tikzpicture}
    };
    \node[below=0pt of output_pic] {\footnotesize Output};
    
    \node[below=of noisy_pic, xshift=1.2cm] (loss) {Loss};
    
    \node[coordinate, below=of output_pic, yshift=0.05cm] (output_to_loss) {};
    \node[coordinate, below=of grad_pic, yshift=0.05cm] (grad_to_loss) {};
    
    \draw[arrow] (image_pic) -- (grad_pic);
    \draw[arrow] (grad_pic) -- (noisy_pic);
    \draw[arrow] (noisy_pic) -- (network);
    \draw[arrow] (network) -- (output_pic);
    \draw[arrow] (output_to_loss) |- (loss.east);
    \draw[arrow] (grad_to_loss) |- (loss.west);
    
    \end{tikzpicture}
    \caption{Training pipeline for an image gradient prior. The image gradient is computed from the image, additive white gaussian noise is added with a randomly selected noise level, the noisy gradient and noise level is passed through the network, and the output is compared to the clean image gradient via the loss.}
    \label{fig:training pipeline}
\end{figure*}

Historically, an important direction for the development of image priors was the decomposition of images into transformed features, such as the image gradient~\cite{rudin.etal1992} or wavelets~\cite{mallat2009}.
While such decompositions played a central role in classical signal processing and image reconstruction, they have been largely underutilized in modern deep learning.
An important framework in signal processing for incorporating these decompositions in image reconstruction is that of \textit{analysis priors}\cite{elad.etal2006, selesnick.etal2009}.
In this problem formulation, priors impose some structure, commonly sparsity, on the transformed representation of the reconstructed image.
Despite their importance in classical signal processing, analysis priors have received little attention in a deep learning context.
While deep learning methods are highly effective in the image domain, 
the success of classical decompositions motivates their re-examination in modern deep-learning settings.

In this work,
we develop a learned image gradient prior which is incorporated within an Analysis PnP Half Quadratic Splitting \textit{(APnP-HQS)} reconstruction algorithm and an Analysis PnP Alternating Direction Method of Multipliers \textit{(APnP-ADMM)} reconstruction algorithm.
We experimentally demonstrate the viability of such priors by showing that they can achieve similar reconstruction performance to those obtained with standard learned image priors. 
These results demonstrate the potential of modern deep analysis priors, such as learned image gradient priors, when used within reconstruction algorithms for image restoration.

\section{Background}
\label{sec:background}
\subsection{Imaging Inverse Problems}
\noindent
Many imaging systems can be formulated as
\begin{equation}
    \ybm = \Abm \xbm + \nbm,
\end{equation}
where $\Abm \in \R^{m\times n}$ is the forward model of the imaging system, $\xbm\in\R^n$ denotes the  image to recover, $\nbm\in\R^m$ represents noise, and $\ybm\in\R^m$ is the measurement.
The goal is to solve an \textit{inverse problem}, where $\xbm$ is recovered from $\ybm$. 
It is common to formulate the inverse problem as an optimization 
\begin{equation} \label{eq: optimization problem}
    \argmin_{\xbm} \ f(\xbm) + g(\xbm),
\end{equation}
where $f$ is the \textit{data-fidelity} term enforcing measurement consistency and $g$ is the \textit{prior} or \textit{regularization} term that restricts the feasible solutions of $\xbm$.
One of the most commonly used $f$ is $\frac{1}{2} \| \Abm \xbm - \ybm \|_2^2$, which arises from the maximum likelihood estimation under the assumption of additive white Gaussian noise in $\ybm$~\cite{boyd.vandenberghe.2004}.
The prior term $g$ is meant to enforce desired features for the reconstructions.
A classical example is that of total variation (TV)\cite{rudin.etal1992}, which enforces the piecewise smoothness of images.

\subsection{Analysis Signal Priors}
\noindent
The \textit{analysis formulation} of Eq.~\eqref{eq: optimization problem} places a prior on a transformed version of $\xbm$~\cite{elad.etal2006, selesnick.etal2009}.
This transformation has historically been a linear map.
Letting the linear transformation be $\Dbm$, Eq.~\eqref{eq: optimization problem} can be written with an analysis prior as 
\begin{equation} \label{eq: analysis optimization problem}
    \argmin_{\xbm} \ f(\xbm) + g( \Dbm \xbm).
\end{equation}
Popular versions of $\Dbm$ are the image gradient and wavelet transform~\cite{rudin.etal1992, mallat2009}, and $g$ is commonly a sparsity-enforcing norm, such as the $\ell_1$ norm.
An example is anistropic TV, where $\Dbm$ is the image gradient and $g$ is the $\ell_1$ norm:
$\| \Dbm \xbm \|_1$. 
The image gradient is the finite difference of the image $\xbm$ along the horizontal and vertical axes.
This can be written as the convolution along the horizontal and vertical axes of the image with the kernel $\dbm = [1,  \text{-}1]$.
In this work, we only consider the case when $\Dbm$ is the image gradient.
\begin{figure*}[t] %
\begin{minipage}[t]{0.48\textwidth}
    \begin{algorithm}[H]
         \caption{APnP-HQS}
         \begin{algorithmic}[1]
         \label{alg: analysis hqs}
         \renewcommand{\algorithmicrequire}{\textbf{input:}}
             \REQUIRE forward model $\mathbf{A}$, measurement $\boldsymbol{y}$, gradient denoiser $\gradDen$, image noise level $\sigma$, denoiser prior strength schedule $\sigma_t$, and regularization parameter $\lambda$.
             \STATE Initialize $\xbm_0$ from $\ybm$; let $\alpha_t := \lambda \sigma^2/\sigma_t^2$
            \FOR {$t = 1$ to $T$}
                    \STATE $\zbm_t = \gradDen(\Dbm \xbm_{t-1}, \sigma_t)$
                    \STATE $\xbm_t = \argmin_{\xbm} \| \Abm \xbm -  \ybm\|_2^2 + \alpha_t \| \Dbm \xbm - \zbm_{t} \|_2^2 $
            \ENDFOR
             \RETURN $\xbm_T$ 
         \end{algorithmic}
    \end{algorithm}
\end{minipage}
\hfill
\begin{minipage}[t]{0.48\textwidth}
    \begin{algorithm}[H]
         \caption{APnP-ADMM}
         \begin{algorithmic}[1]
         \label{alg: analysis admm}
         \renewcommand{\algorithmicrequire}{\textbf{input:}}
             \REQUIRE forward model $\mathbf{A}$, measurement $\boldsymbol{y}$, gradient denoiser $\gradDen$, image noise level $\sigma$, denoiser prior strength schedule $\sigma_t$, and regularization parameter $\lambda$.
             \STATE Initialize $\xbm_0$ from $\ybm$; initialize $\sbm_0$ to $\bm{0}$;  let $\alpha_t := \lambda \sigma^2/\sigma_t^2$
                \FOR {$t = 1$ to $T$}
                    \STATE $\zbm_t = \gradDen(\Dbm \xbm_{t-1} - \sbm_{t-1}, \sigma_k)$
                    \STATE $\xbm_t = \argmin_{\xbm} \ \|\Abm \xbm - \ybm \|_2^2 + \alpha_t \| \zbm_t + \sbm_{t-1} - \Dbm \xbm \|_2^2 $
                    \STATE $\sbm_t = \sbm_{t-1} + \zbm_t - \Dbm \xbm_t$
                \ENDFOR
             \RETURN $\xbm_T$ 
         \end{algorithmic}
    \end{algorithm}
\end{minipage}
\hfill
\end{figure*}

\subsection{Plug-and-Play Priors}
\noindent
The \textit{Plug-and-Play (PnP)} framework solves Eq.~\eqref{eq: optimization problem} by incorporating a learned prior with a data-fidelity term~\cite{Kamilov.etal2023}. 
These priors are generally trained to be Gaussian image denoisers, although other paradigms have been explored~\cite{Ulyanov.etal2018, delbracio.etal2023, zhou.etal2024, hu.etal2024, terris.etal2025}.
An important aspect of PnP is that the prior is trained without any knowledge of the inverse problem for which it will be used, i.e. it is  trained purely as an image denoiser.
Therefore, the PnP framework provides a method to solve any imaging inverse problem that can be formulated as Eq.~\eqref{eq: optimization problem} without training a separate neural network for each problem. 

While many PnP algorithms have been proposed, the most relevant to this paper is the algorithm based on the alternating direction method of multipliers (ADMM)~\cite{boyd.etal2011} 
and
DPIR~\cite{zhang.etal2022}, which
uses a learned image denoiser in a half-quadratic splitting (HQS)~\cite{geman.yang1995} algorithm. 

The vast majority of the priors used within PnP algorithms are those on the image space itself.
Absent from the current literature is the development of deep neural network-based analysis priors for image reconstruction -- i.e. neural networks learned to represent traditional transformed domains such as image gradients and then used within PnP algorithms.

\section{Method}
\subsection{Image Gradient Prior}
We take inspiration from the analysis problem formulation and propose incorporating a learned prior on the image gradient rather than the image itself.
In particular, we learn an additive white gaussian noise (AWGN) denoiser $\gradDen$ on the image gradient.
This can be interpreted as a deep learning analogue to the total variation prior.

In order to train the analysis prior network, we apply the image gradient operator $\Dbm$ to an image and add synthetic noise $\nbm$ with variance $\sigma^2$.
We use the $\ell_1$ training loss
\begin{equation}
    \label{eq: training loss}
    \big \| \big ( \gradDen(\Dbm \xbm + \sigma \nbm, \sigma), \Dbm \xbm \big \|_1.
\end{equation}
The input to the network is the noisy image gradient and noise level and its output is the denoised image gradient.
Figure \ref{fig:training pipeline} illustrates the training procedure.

\subsection{Analysis PnP Formulation}
For image reconstruction, we use the following analysis formulation:
\begin{equation} \label{eq: PnP analysis optimization problem}
    \argmin_{\xbm} \ f(\xbm) + \lambda \Rcal( \Dbm \xbm).
\end{equation}
The prior is $g = \lambda \Rcal$, where $\Rcal$ is the implicit regularization function of the denoiser $\gradDen$ and $\lambda$ is a regularization strength parameter.
As is standard in PnP methods, we do not have access to the function $\Rcal$ itself; it is only implicitly available through the gradient prior $\gradDen$.

In order to solve general Eq. \eqref{eq: PnP analysis optimization problem}, we propose an HQS and an ADMM inspired PnP algorithm.
Both of these algorithms take advantage of variable splitting.
The difference of our approach with the current PnP literature is that we use a prior on a transformed image rather than the image itself.
Therefore, we use the variable splitting $\zbm = \Dbm \xbm$ instead of $\zbm = \xbm$.
This gives the constrained optimization problem
\begin{equation} \label{eq: constrained optimization problem}
    \argmin_{\xbm} \  \frac{1}{2 \sigma^2} \| \Abm \xbm - \ybm \|_2^2 + \lambda \Rcal(\zbm)  \quad \text{such that} \quad \zbm = \Dbm \xbm.
\end{equation}

\subsection{Analysis PnP-HQS}
To apply HQS to Eq. \eqref{eq: constrained optimization problem}, we reformulate Eq. \eqref{eq: constrained optimization problem} into the following unconstrained optimization problem
\begin{equation} \label{eq: HQS optimization problem}
     \argmin_{\xbm, \zbm }\frac{1}{2\sigma^2}\| \Abm \xbm - \ybm \|_2^2 + \lambda \Rcal(\zbm)  + \frac{\mu}{2} \| \Dbm \xbm - \zbm \|_2^2.
\end{equation}
HQS alternates between minimizing $\xbm$ and $\zbm$, and so each iteration requires solving
\begin{align}
     \zbm_t &= \argmin_{\zbm} \ \lambda \Rcal(\zbm)  + \frac{\mu}{2} \| \Dbm \xbm_{t-1} - \zbm \|_2^2 \label{eq: HQS prior prox}\\
     \xbm_t &= \argmin_{\xbm} \ \frac{1}{2 \sigma^2} \| \Abm \xbm - \ybm \|_2^2  + \frac{\mu}{2} \| \Dbm \xbm - \zbm_t \|_2^2. \label{eq: HQS fidelity optimization}
\end{align}
As is the standard approach in the PnP framework, line \eqref{eq: HQS prior prox} can be replaced by 
\begin{equation*}
    \gradDen\big(\Dbm \xbm, \sqrt{\lambda /\mu}\big) \;.
\end{equation*}
In order for HQS to converge, $\mu$ must be large, which requires a large number of iterations $T$.
Therefore, following \cite{zhang.etal2017}, we use a schedule of increasing $\mu_t$.
Since $\mu$ controls the prior strength for $\gradDen$, this is equivalent to a decreasing noise schedule $\sigma_t$.
For a complete discussion of this schedule, we refer to~\cite{zhang.etal2017, zhang.etal2022}.
The full APnP-HQS algorithm is shown in Algorithm \ref{alg: analysis hqs}.

\setlength{\tabcolsep}{12pt}
\renewcommand{\arraystretch}{1.1}
\begin{table*}[tbp]
    \centering
    \caption{
        Performance for image deblurring and super resolution comparing image-space PnP with gradient space Analysis PnP (APnP).
        A scale factor (SF) of $1$ indicates deblurring, while scale factors $2$ and $3$ are for $2\times$ and $3 \times$ super resolution.
        We report performance for noiseless and noisy measurements.
        Each scale factor/noise result is averaged over $8$ blur kernels.
    }
    \label{table: results}
    \vspace{0.2em}
    \begin{minipage}{0.47\textwidth}
    \centering
    \textbf{SSIM} \\
    \vspace{0.5em}
    \setlength{\tabcolsep}{6pt} %
        \resizebox{\columnwidth}{!}{\begin{tabular}{ c c c c c c }
            \hline
            \textbf{SF} & \textbf{Noise} & \textbf{DPIR} & \textbf{APnP-HQS} & \textbf{PnP-ADMM} & \textbf{APnP-ADMM} \\
            \hline
            \multirow{2}{*}{1}
              & 0    & 0.8746 & 0.8711 & 0.8618 & 0.8541 \\
              & 7.65 & 0.7447 & 0.7210 & 0.7297 & 0.7282 \\
            \hline
            \multirow{2}{*}{2}
              & 0    & 0.8385 & 0.8371 & 0.8329 & 0.8287 \\
              & 7.65 & 0.6912 & 0.6689 & 0.6786 & 0.6731 \\
            \hline
            \multirow{2}{*}{3}
              & 0    & 0.7754 & 0.7740 & 0.7732 & 0.7691 \\
              & 7.65 & 0.6501 & 0.6330 & 0.6416 & 0.6321 \\
            \hline
        \end{tabular}}
    \end{minipage}
    \hfill
    \begin{minipage}{0.47\textwidth}
        \centering
        \textbf{PSNR} \\
        \vspace{0.5em}
        \setlength{\tabcolsep}{6pt} %
        \resizebox{\columnwidth}{!}{
        \begin{tabular}{c c c c c c}
            \hline
            \textbf{SF} & \textbf{Noise} & \textbf{DPIR} & \textbf{APnP-HQS} & \textbf{PnP-ADMM} & \textbf{APnP-ADMM} \\
            \hline
            \multirow{2}{*}{1}
              & 0    & 31.67 & 31.40 & 31.11 & 30.79 \\
              & 7.65 & 27.06 & 26.72 & 26.66 & 26.58 \\
            \hline
            \multirow{2}{*}{2}
              & 0    & 28.91 & 28.81 & 28.69 & 28.50 \\
              & 7.65 & 25.77 & 25.53 & 25.50 & 25.41 \\
            \hline
            \multirow{2}{*}{3}
              & 0    & 27.13 & 27.02 & 26.98 & 26.85 \\
              & 7.65 & 24.91 & 24.74 & 24.69 & 24.59 \\
            \hline
        \end{tabular}
        }
    \end{minipage}%
\end{table*}

\subsection{Analysis PnP-ADMM}
For the ADMM variant (APnP-ADMM), we use the following augmented Lagrangian for Eq.~\eqref{eq: constrained optimization problem}
\begin{equation} \label{eq: ADMM optimization problem}
     \frac{1}{2 \sigma^2} \| \Abm \xbm - \ybm \|_2^2 + \lambda \Rcal(\zbm)  + \mu \sbm^{\Tsf}(\zbm - \Dbm \xbm) + \frac{\mu}{2} \| \zbm - \Dbm \xbm \|_2^2 \;.
\end{equation}
Each iteration of ADMM requires solving
\begin{align}
     \zbm_t &= \argmin_{\zbm} \ \lambda \Rcal(\zbm) + \frac{\mu}{2}\| \zbm - (\Dbm \xbm_{t-1} - \sbm_{t-1}) \|_2^2 \label{eq: ADMM prior prox}\\
     \xbm_t &= \argmin_{\xbm} \ \frac{1}{2 \sigma^2} \|\Abm \xbm - \ybm \|_2^2 \nonumber \\
        & \qquad \qquad + \frac{\mu}{2} \| (\zbm_t + \sbm_{t-1}) - \Dbm \xbm \|_2^2 \label{eq: ADMM fidelity optimization} \\
     \sbm_t &= \sbm_{t-1} + \zbm_t - \Dbm \xbm_t \;.
\end{align}
Line \eqref{eq: ADMM prior prox} is again replaced by $\gradDen$ and the entire APnP-ADMM algorithm is displayed in Algorithm \ref{alg: analysis admm}.
We adopt the same noise schedule as described in APnP-HQS.

\begin{figure*}[tp]
 \centering
 \includegraphics[width=0.99\textwidth]{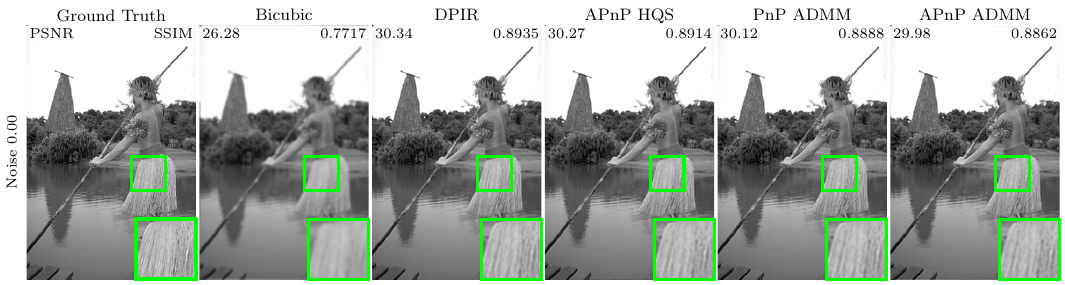} %
 \caption{Visual reconstruction example of the four reconstruction algorithms for the $2 \times$ image super resolution task.
 }
\label{fig: visual result}
\end{figure*}
\section{Experiments}

\subsection{Network Training}
Both the image-space and gradient-space denoisers follow the DRUNet architecture~\cite{zhang.etal2022}.
In both cases, there is an extra channel for the noise level map so that the network is aware of the noise standard deviation.
Therefore, the image-space network has an input of $2$ channels and an output of $1$ channel and the gradient-space network has an input of $3$ channels and an output of $2$ channels.
We use the trained grayscale denoiser from~\cite{zhang.etal2020} and trained our own image-gradient denoiser using the same natural image dataset.
The images are normalized to $[0,1]$ by dividing each image by $255$.
For training, we minimize Eq. \eqref{eq: training loss} with the Adam optimizer~\cite{kingma.adam2015}.
We use a batch size of $128$ and crop random $128\times 128$ patches from the images.
For each patch $\xbm$, the input is $\Dbm \xbm + \sigma \nbm$ where $\nbm \sim \Ncal(0,\Ibm)$ and $\sigma \sim \text{Uniform}(0, 71/255)$.
We chose this range because image-space DRUNet uses $\sigma \sim \text{Uniform}(0, 50/255)$ and taking the image gradient of noise with variance $\sigma^2$ results in noise with variance $2 \sigma^2$.

\subsection{Image Reconstruction} 
We test the trained priors within four PnP algorithms to solve the image deblurring and image super resolution tasks.
We report results for the anlysis algorithms APnP-HQS and APnP-ADMM along with their non-analysis forms DPIR and PnP-ADMM. 
Note that DPIR is a PnP-HQS algorithm.
Following DPIR, we use $24$ iterations for all algorithms and adopt the same noise schedule $\sigma_t$.

The forward model of the image super resolution is 
\begin{equation*}
    \ybm = (\xbm \circledast \kbm)\downarrow_{s} + \nbm,
\end{equation*}
where $\kbm$ is the blurring kernel and $\downarrow_{s}$ denotes downsampling by $s$.
For image deblurring, $s=1$.
Following~\cite{zhang.etal2020}, we use $8$ blur kernels, $4$ of which are isotropic and the rest anisotropic.
The fast implementation from~\cite{zhao.etal2016} is used to solve Eq.~\eqref{eq: HQS fidelity optimization} and Eq.~\eqref{eq: ADMM fidelity optimization} in APnP-HQS and APnP-ADMM.

Following DPIR, we use $24$ iterations, adopting the same noise schedule $\sigma_t$; however, for APnP-HQS we found multiplying by $\sqrt{2}$ resulted in better performance.
As reported in \cite{zhang.etal2022}, we set $\lambda=0.23$ for DPIR.
We empirically found $\lambda=0.18$ to give the best results for APnP-HQS.
We used the same number of iterations and noise schedule for PnP-ADMM and APnP-ADMM.
We empirically set $\lambda=0.38$ for PnP-ADMM and $\lambda = 0.24$ for APnP-ADMM.

The $68$ images from the BSD68 dataset~\cite{martin.etal2001} are used for the reconstruction experiments.
As can be seen numerically and visually, the analysis versions of PnP achieve comparable performance to their standard PnP counterparts.
While we generally do not outperform standard PnP, we believe our results encourage further exploration into developing analysis-based priors and architectures more suited for them.
In this work we have only considered the image gradient, but other transformations, such as the wavelet transform, could result in improved analysis-based PnP algorithms.

\section{Conclusion}
In this work, we propose using an image gradient denoiser as a prior within an HQS and ADMM PnP algorithm, which we call APnP-HQS and APnP-ADMM, respectively.
Empirical results show that APnP-HQS and APnP-ADMM can obtain comparable reconstructions to  standard PnP.
While we do not achieve state-of-the-art, this work demonstrates that analysis priors can be extended for the deep learning era. We believe further work on the variable splitting and inference algorithms can close this gap.

{
\small

\bibliographystyle{IEEEbib}
\bibliography{refs}
}

\end{document}